\documentclass[twocolumn,tighten,times]{aastex62}
\usepackage{rotating,ulem}
\newcommand{\err}[2]{\ensuremath{^{_{+#1}}_{^{-#2}}}}
\newcommand{\ee}[2]{\ensuremath{{#1}\!\times\!10^{#2}}}
\newcommand{\msun}{\ensuremath{\mathrm{M}_\odot}}
\newcommand{\ergsec}{\ensuremath{\mathrm{erg~s}^{-1}}}
\newcommand{\ergcms}{\ensuremath{\mathrm{erg~cm}^{-2}~\mathrm{s}^{-1}}}
\newcommand{\Lx}{\ensuremath{L_\mathrm{x}}}
\newcommand{\Fx}{\ensuremath{F_\mathrm{x}}}

\newcommand{\Ha}{\ensuremath{\mathrm{H\alpha}}}
\newcommand{\pcmsq}{\ensuremath{\mathrm{cm}^{-2}}}
\newcommand{\gcc}{\ensuremath{\mathrm{g\ cm}^{-3}}}
\newcommand{\xmm}{\textit{XMM}}
\newcommand{\xmmn}{\textit{XMM-Newton Observatory}}
\newcommand{\swift}{\textit{Swift}}
\newcommand{\chandra}{\textit{Chandra}}
\newcommand{\cxo}{\textit{Chandra X-ray Observatory}}
\newcommand{\kms}{\ensuremath{\mathrm{km~s}^{-1}}}

\newcommand{\dave}[1]{{\color{black}  #1}}
\newcommand{\revision}[1]{{\color{red} #1}}

\received{2018 March 23}
\accepted{2019 August 22}
\published{2019 September 26}
\submitjournal{ApJ}

\shorttitle{SN 2004dk Late-time Interaction} 
\shortauthors{Pooley et al.}
\begin{document}

\title{Interaction of SN~Ib 2004dk with a Previously-Expelled Envelope}

\author{David Pooley}
\affiliation{Department of Physics and Astronomy, Trinity University, San Antonio, TX, USA}
\affiliation{Eureka Scientific, Inc., USA}

\author{J.\ Craig Wheeler}
\affiliation{Department of Astronomy, University of Texas at Austin, Austin, TX, USA}

\author{Jozsef Vink{\'o}}
\affiliation{Department of Astronomy, University of Texas at Austin, Austin, TX, USA}
\affiliation{Konkoly Observatory, Research Centre for Astronomy and Earth Sciences, Hungarian Academy of Sciences, H-1121 Budapest, Hungary}
\affiliation{Department of Optics and Quantum Electronics, University of Szeged, D\'om t\'er 9, Szeged, 6720 Hungary}

\author{Vikram V. Dwarkadas}
\affiliation{Department of Astronomy and Astrophysics, University of Chicago, Chicago, IL, USA}

\author{Tamas Szalai}
\affiliation{Konkoly Observatory, Research Centre for Astronomy and Earth Sciences, Hungarian Academy of Sciences, H-1121 Budapest, Hungary}
\affiliation{Department of Optics and Quantum Electronics, University of Szeged, D\'om t\'er 9, Szeged, 6720 Hungary}

\author{Jeffrey M.\ Silverman}
\affiliation{Department of Astronomy, University of Texas at Austin, Austin, TX, USA}
\affiliation{Samba TV, San Francisco, CA, USA}

\author{Madelaine Griesel}
\affiliation{Department of Physics and Astronomy, Trinity University, San Antonio, TX, USA}

\author{Molly McCullough}
\affiliation{Department of Physics and Astronomy, Trinity University, San Antonio, TX, USA}

\author{G.\ H.\ Marion}
\affiliation{Department of Astronomy, University of Texas at Austin, Austin, TX, USA}

\author{Phillip MacQueen}
\affiliation{Department of Astronomy, University of Texas at Austin, Austin, TX, USA}

\correspondingauthor{David Pooley}
\email{dpooley@trinity.edu}

\begin{abstract}

The interaction between the expanding supernova (SN) ejecta with the circumstellar material (CSM) that was expelled from the progenitor prior to explosion is a long-sought phenomenon, yet observational evidence is scarce. Here we confirm a new example: SN~2004dk, originally a hydrogen-poor, helium-rich Type~Ib SN that reappeared as a strong \Ha-emitting point-source on narrowband \Ha\ images. We present follow-up optical spectroscopy that reveals the presence of a broad \Ha\ component with full width at half maximum of $\sim$\,290 \kms\ in addition to the narrow  \Ha +[\ion{N}{2}] emission features from the host galaxy. Such a broad component is a clear sign of an ejecta-CSM interaction. We also present observations with the \xmmn, the \swift\ satellite, and the \cxo\ that span 10 days to 15 years after discovery. The detection of strong radio, X-ray, and \Ha\ emission years after explosion allows various constraints to be put on pre-SN mass-loss processes.  We present a wind-bubble model in which the CSM is ``pre-prepared" by a fast wind interacting with a slow wind. Much of the outer density profile into which the SN explodes corresponds to no steady-state mass-loss process. We estimate that the shell of compressed slow wind material was ejected $\sim$1400 yr prior to explosion, perhaps during carbon burning, and that the SN shock had swept up about 0.04 \msun\ of material. The region emitting the \Ha\ has a density of order $10^{-20}\, \gcc$. 
     
\end{abstract}

\section{Introduction}

One of the major issues in supernova (SN) research is how stars that begin with $\sim$ 70\% hydrogen on the main sequence are nearly or completely devoid of hydrogen by the time they explode. The hydrogen is lost by steady winds or bouts of mass loss or complex processes of binary star interaction involving one or more stages of common envelope evolution. In many proposed evolution scenarios some of the expelled hydrogen is expected to remain in a circumstellar medium (CSM). After the SN explodes, it will eventually run into that expelled matter after months, years, or even decades. Extensive mass loss has been associated with both Wolf-Rayet stars and luminous blue variables \citep[LBVs;][]{smith14}. Studying late-time interaction provides a means of probing the transition to and the duration of the Wolf-Rayet phase in massive stars and gaining a deeper understanding of the LBVs and other mass-loss phenomena. 

There are several ways to study the late-time interaction of stripped-envelope supernovae (SNe) with previously-shed material \citep[for a review of interaction emission, see][]{2017hsn..book..875C}.  X-ray and radio emission are very likely signs of interaction and can provide mass-loss diagnostics.  In the optical, the appearance of \Ha\ emission at late times would be another sure sign of interaction; however, without a baseline measure of \Ha\ flux in the vicinity of an SN at early times, the detection of \Ha\ at late times could be ambiguous, as nearby \ion{H}{2} regions would also emit \Ha.  Associating detected \Ha\ with interaction could be based on the following, in order of increasing level of conviction: (1) consistency with a point source; (2) \Ha\ luminosity above that of typical \ion{H}{2} regions ($\sim$10$^{37}$--10$^{39}$); (3) variability; or (4) broadening of the \Ha\ line beyond that expected for an \ion{H}{2} region.

The first example of a hydrogen-deficient SN undergoing late-time collision with previously expelled envelope material was the Type Ib/c SN~2001em \citep{2006ApJ...641.1051C}, characterized by late-time radio \citep{2004IAUC.8282....2S}, X-ray \citep{2004IAUC.8323....2P}, and H$\alpha$ emission \citep{2004GCN..2586....1S}.  Late-time interaction was also seen in the Type~Ia PTF~11kx \citep{Dilday12,Silverman13a,Graham17}.  

Motivated by the example of 2001em, we began searching for late-time interaction in 2009 with a Cycle 11 \cxo\ proposal (PI: Pooley) targeting five old Type Ib/c SNe.  Additionally, in order to increase the sample size substantially, we have been using the DIAFI imager on the Harlan J.\ Smith 2.7 m telescope at McDonald Observatory since 2014 February to search for evidence of delayed collision and excitation of \Ha\ with narrowband filters, one near the expected redshifted wavelength of \Ha\ and another in an ``off" band for calibration. Starting with a list of 3662 SNe, we have selected SN~Ia, and SN~IIb, SN~Ib, and SN~Ic core-collapse SNe that are closer than $\sim$35 Mpc and with decl.\ $\gtrsim -30$. This selection produced a list of 178 target SNe. \citet{Vinko17} summarized our early results. With our narrowband imaging we have found evidence of \Ha\ at the location of the SNe for 13 of 99 events for which we have obtained and fully reduced the data through 2016. We confirmed the broad \Ha\ in SN~2014C, previously reported by \citet[][see also \citealt{Anderson17,Margutti17}]{Milisavljevic15}. 

In our multi-epoch, narrowband imaging, we have found strong evidence of variability in three of 38 SNe observed more than once.  Variability is sufficient, but not necessary, evidence of interaction. The CSM shock could be propagating through an extended medium generating roughly constant \Ha\ flux as in the SN~IIn 1988Z \citep{88Z}.  The strength of the H$\alpha$ line is also an important clue if it is substantially larger than expected from an \ion{H}{2} region. An extreme example is SN~2005kl, which showed $L_\Ha \sim 7\times10^{41}$ \ergsec. Such a high \Ha\ luminosity strongly suggests the presence of CSM interaction. The most compelling \Ha\ evidence of interaction is the detection of the broadened \Ha\ emission. 

Here we present optical spectra and X-ray observations of SN~2004dk that confirm this event to belong to the class of late-interacting SNe.  Section \ref{04dk} describes our observations of SN~2004dk. Section \ref{discuss} presents some implications of our results. Conclusions and a look forward are given in \S\ref{concl}.

\section{SN~2004dk Observations and Analysis}
\label{04dk}

\subsection{Detection of \Ha}
\label{halphaobs}

SN~2004dk (a helium-rich, hydrogen-poor Type Ib) was detected in X-rays shortly after explosion \citep{Pooley07} and by radio observations at later phases \citep{Stockdale09, Wellons12}. Early nebular spectra showed only a weak, narrow, unresolved H$\alpha$ emission feature \citep{Maeda08,Shivvers17}. Our narrowband imaging on 2014 February 27,
2015 March 14, and 2016 June 9 provided a strong (signal-to-noise ratio $>$ 50) detection of a point source at the position of the SN that suggested that the blast wave had finally reached the expelled H-rich envelope. SN~2004dk had an \Ha\ luminosity of $1.5 \times 10^{39}$ \ergsec, which is similar to that of SN~2001em ($\sim 10^{39}$ \ergsec). The lack of distinct H$\alpha$ variability in our narrowband imaging left some doubt \citep{Vinko17}. Our spectra taken in the spring of 2017 removed that doubt. In response to \citet{Vinko17}, \citet{mauerhan18} obtained and reported a broadened H$\alpha$ profile of SN~2004dk. Here we present our narrowband imaging and spectral data and address the broader context in which they fit.

The late-time spectra on SN 2004dk were obtained with the Low Resolution Spectrograph 2 (LRS2) on the 10m Hobby-Eberly Telescope (HET). LRS2 consists of two dual-arm spectrographs, covering 370--470 nm (LRS2-B UV), 460--700 nm (LRS2-B Orange), 650--842 nm (LRS2-R Red), and 818--1050 nm (LRS2-R Far-Red), with an average spectral resolution of $R \sim$ 1500 \citep{2016SPIE.9908E..4CC}. 

\begin{figure}%[!ht]
\includegraphics[width=\columnwidth]{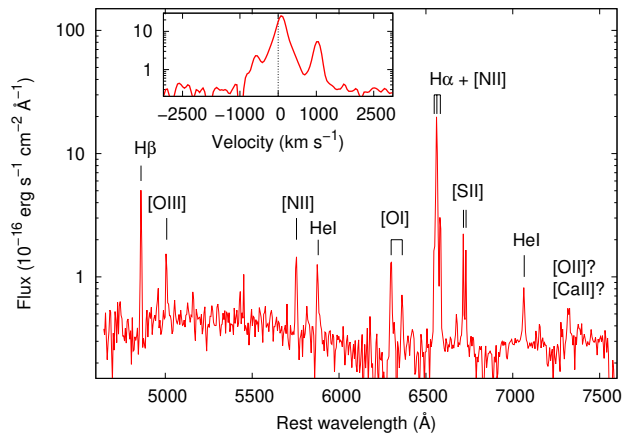}
\caption{Combined LRS2 spectra of SN~2004dk showing the strong, broadened \Ha\ and also H$\beta$, [\ion{O}{1}], and \ion{He}{1}. The latter are only seen in SN~2004dk and SN~2014C in our sample and give further support that a SN/CSM interaction is occurring in SN~2004dk 13 yr after the explosion. The inset shows the same spectrum in velocity space in the vicinity of \Ha.}
\label{full_04dk}
\end{figure}

For SN~2004dk, two spectra, taken on 2017 March 30 and 31 were acquired with the Red arm of LRS2-R, while a third spectrum was obtained with the Orange arm of LRS2-B on 2017 May 24. Figure \ref{full_04dk} presents the full, combined LRS2-B + LRS2-R spectrum \dave{from 2017}. The inset shows a blowup of the region of \Ha\ in velocity space.  \dave{We obtained another spectrum on 2019 June 21 which was virtually identical.}

\subsection{Spectral Properties}
\label{specprop}
 
Table~\ref{tab:lineid} shows the list of features that are identified in the combined HET spectrum, based on the line identifications by \citet{Milisavljevic15} for the interacting SN~2014C. It is seen that SN~2004dk and SN~2014C have many emission features in common: beside the strong hydrogen features, lines of \ion{He}{1}, [\ion{O}{1}], [\ion{O}{2}], [\ion{N}{2}], and [\ion{S}{2}] are present in the spectrum of both SNe. SN~2004dk lacks the broad \ion{Ca}{2} features that were strong in the spectrum of SN~2014C \citep{Milisavljevic15}; in contrast, the \ion{He}{1} features appear stronger in SN~2004dk. This could be attributed to somewhat different ejecta composition of the two events, or it could be due to the timing of the interaction.

The full widths at half maximum (FWHMs) of the identified features (measured by Gaussian fitting) are also listed in Table~\ref{tab:lineid}. It is expected that the features that are associated with the expanding ejecta and the interacting region should be somewhat broader than those features that originate from the interstellar medium (ISM). Indeed, \Ha\, the \ion{He}{1} features, the [\ion{N}{2}] $\lambda$5755 and the [\ion{O}{1}] $\lambda \lambda$6300,6364 features appear somewhat more broadened than the other lines; however, all of these features are just barely resolved (the instrumental resolution being $\sim 4$ \AA), thus, their FWHMs do not fully reflect their intrinsic line width (except for \Ha\ and maybe \ion{He}{1}; see below).

\begin{deluxetable}{lccc}
\tablecaption{Emission line identification for SN 2004dk \label{tab:lineid}}
\tablehead{
\colhead{Ion/Feature} & \colhead{$\lambda$} & \colhead{FWHM\tablenotemark{a}} & \colhead{FWHM$_{i}$\tablenotemark{b}} \\
  & (\AA) & (\AA) & (\AA) 
}
\startdata
H$\beta$ & 4861 & 5.32 & 3.24\\
{[\ion{O}{3}]} & 4959 & 5.12 & 3.31\\
{[\ion{O}{3}]} & 5007 & 5.14 & 3.34 \\
{[\ion{N}{2}]} & 5755 & 8.40 & 3.83\\
{\ion{He}{1}} & 5876 & 8.45 & 3.92\\
{[\ion{O}{1}]} & 6300 & 7.84 & 4.20\\
{[\ion{O}{1}]} & 6364 & 9.20 & 4.24\\
{[\ion{N}{2}]} & 6548 & 3.50 & 4.36\\
\Ha & 6563 & 6.55 & 4.37\\
{[\ion{N}{2}]} & 6583 & 4.42 & 4.39\\
{[\ion{S}{2}]} & 6716 & 4.02 & 4.48\\
{[\ion{S}{2}]} & 6731 & 4.21 & 4.49\\
{\ion{He}{1}} & 7066 & 9.64 & 3.92 
\enddata
\tablenotetext{a}{Measured by Gaussian fitting.}
\tablenotetext{b}{Computed from instrumental resolution.}
\end{deluxetable}

The flux ratios of forbidden lines of [\ion{N}{2}] and [\ion{S}{2}] can be used to estimate the electron density and temperature from the nebular-phase spectra \citep[e.g.][]{1974agn..book.....O}.  We measure the line flux ratio of $(I(\lambda 6548)+I(\lambda 6583))/I(\lambda 5755) \approx 3.35 \pm 0.2$ for [\ion{N}{2}]. Following \citet{Milisavljevic15}, we apply the {\tt TEMDEN} task in IRAF to get a lower limit for the electron density of $\gtrsim 2 \times 10^{5}$ cm$^{-3}$, corresponding to $T_e \lesssim 3.3 \times 10^4$ K. A temperature of $T_e \sim 10^4$ K would imply $N_e \sim 7 \times 10^5$ cm$^{-3}$. On the contrary, the measured flux ratio of $I(\lambda 6716) /I(\lambda 6731) \approx 1.40 \pm 0.1$ for [\ion{S}{2}] would require $N_e \sim 22$ cm$^{-3}$ if the temperature were $T_e \sim 10^4$ K, which is similar to the densities of the outer regions in planetary nebulae \citep[e.g][]{1974agn..book.....O}. It is not possible to get a self-consistent solution from these two diagnostic line ratios, as the [\ion{S}{2}] ratio implies $T_e < 1.4 \times 10^4$ K. This suggests that the formation of these features takes place at different locations along the line of sight: the [\ion{N}{2}] features that correspond to higher electron densities might be formed close to the shock-compressed circumstellar medium (CSM), see \S\ref{halphamod}, while the [\ion{S}{2}] features more likely originate from the unshocked interstellar medium (ISM), far from the SN.

\subsection{Analysis of the \Ha\ profile}
\label{profile}

\begin{figure}[htp]
\includegraphics[width=\columnwidth]{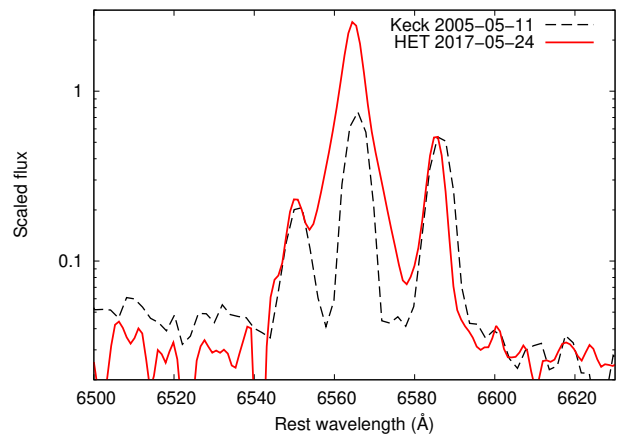}
\caption{The spectrum of SN~2004dk obtained with HET LRS2 on 2017 March 30 compared to a much earlier spectrum obtained with Keck on 2005 May 11 \citep{Shivvers17}. The \Ha\ line is flanked by the two lines of [\ion{N}{2}] at 6548\AA\ and 6583\AA. The Keck spectrum shows only the narrow line corresponding to a nearby \ion{H}{2} region; the HET spectrum reveals clear broadening of the \Ha\ line, showing that ejecta/CSM interaction had begun.}
\label{hal_04dk}
\end{figure}

Figure~\ref{hal_04dk} shows the \Ha\ profile of our 2017 spectrum in contrast to a 2005 Keck spectrum \citep{Shivvers17}, downloaded from the Weizmann Interactive Supernova Data Repository\footnote{https://wiserep.weizmann.ac.il/} and scaled to match the strengths of the [\ion{N}{2}] 6548\AA\ and 6583\AA\ features of the new late-time spectrum. The Keck spectrum,  taken with the Low Resolution Imaging Spectrograph (LRIS) with a resolution of $\sim 2000$ that corresponds to $\sim 150$ km~s$^{-1}$ around \Ha, shows slight broadening beyond instrumental. In our later spectra, the \Ha\ line shows a FWHM width of $\sim$290 \kms (see below). The line wings nearly overlap with the nearby lines of [\ion{N}{2}] at 6548\AA\ and 6583\AA. The collision is ensuing. 

\begin{figure}%[!ht]
\includegraphics[width=\columnwidth]{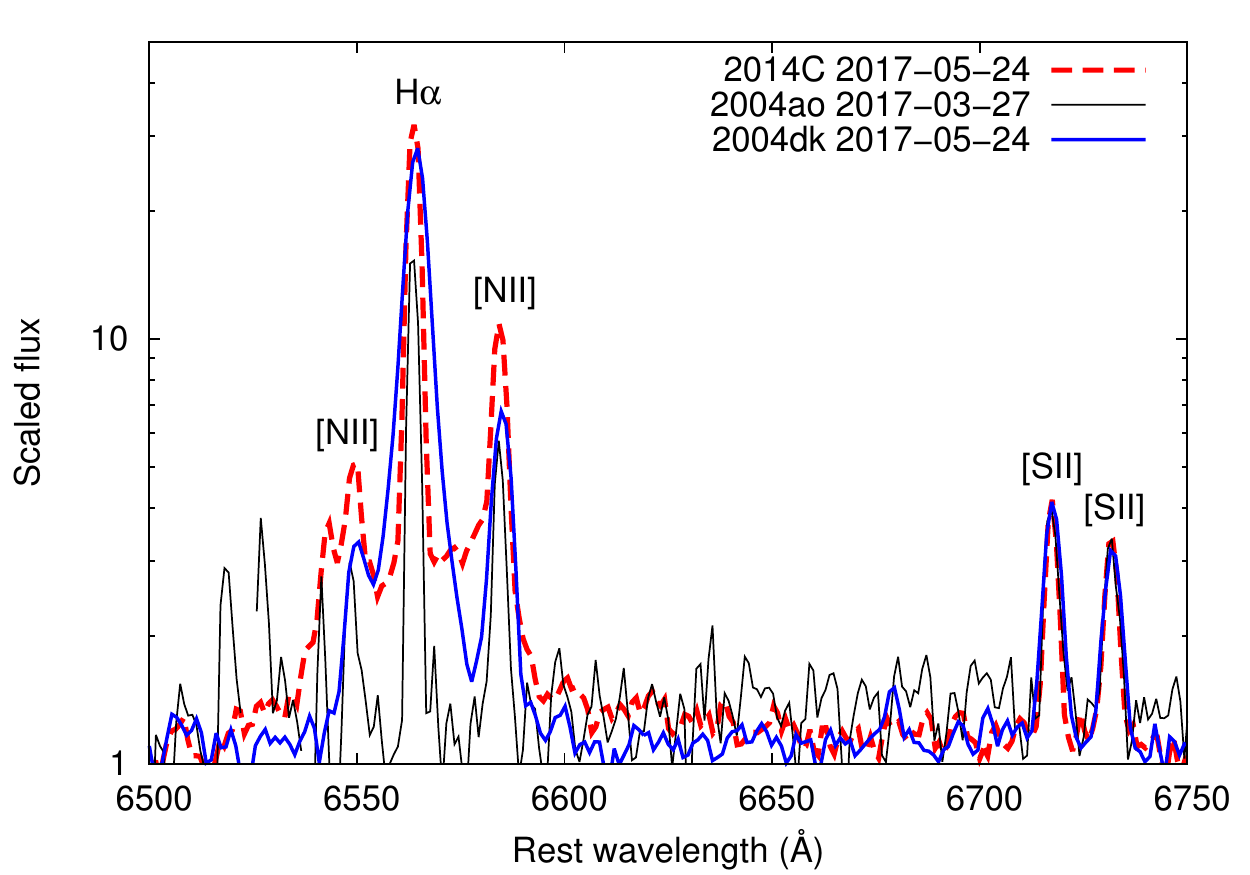}
\caption{The \Ha\ profiles of three SNe observed in our program. SN~2004ao (green) shows only instrumental broadening. SN~2004dk (blue) shows an intermediate level of broadening, distinctly more than instrumental. SN~2014C (red) shows both a central narrow, unresolved \Ha\ line and a substantially broader component. The two lines to the right are [SII].}
\label{compare}
\end{figure}

Figure \ref{compare} gives LRS2 spectra near \Ha\ for three events. SN~2004ao shows \Ha\ in emission, but no signs of variability or broadening. SN~2014C, the youngest late-time interaction, shows both a narrow central unresolved component and a broad component, extending even beyond the [\ion{N}{2}] lines. SN~2004dk shows a definite but more modest broadening of the \Ha\ line, which may be expected for interaction happening in an older SN.

\begin{figure}%[!ht]
\includegraphics[width=\columnwidth]{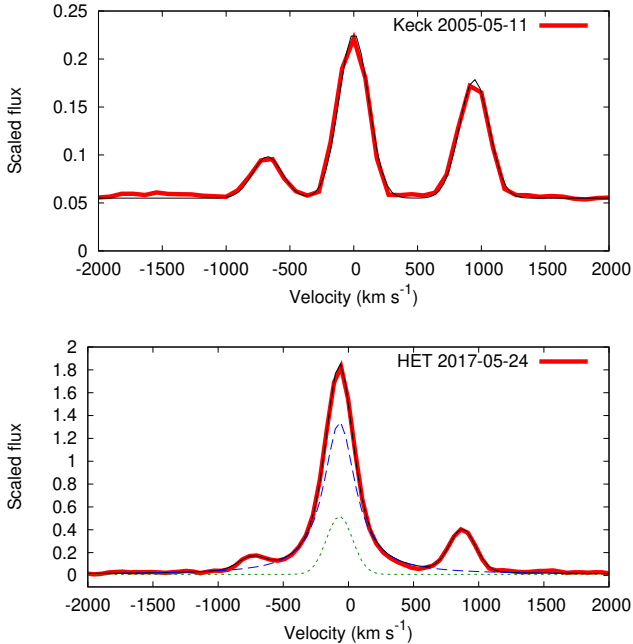}
\caption{Top panel: fitting the 2005 Keck spectrum (red) with three Gaussians (black). Bottom panel: fitting the 2017 HET spectrum with three Gaussians plus a Lorentzian. The Gaussian and the Lorentzian components of \Ha\ are plotted with green dotted and blue dashed lines, respectively. See text for more details on the fitting.}
\label{fwhmfit}
\end{figure}

\citet{mauerhan18} estimated the FWHM of \Ha\ from their Keck spectrum taken on 2017 May 29, nearly contemporaneously with our LRS2-B spectrum. They assumed that the \Ha\ line profile consists of a narrow (FWHM $\sim$130 \kms) Gaussian and a broader (FWHM $\sim$400 \kms) Lorentzian component with the narrow component having a factor of $\sim$2 higher amplitude than the Lorentzian one. Since the FWHM is affected by the amplitude ratio of the components, here we revisit this issue by refitting the \Ha\ line profile with a similar model; however, instead of fixing the amplitude of the Gaussian component or letting it float, we first fit the \Ha\ and the [\ion{N}{2}] $\lambda \lambda$ 6548,6583 features in the 2005 May 11 Keck spectrum with three Gaussians (Fig.~\ref{fwhmfit} top panel), then use the same amplitude ratio for \Ha/[\ion{N}{2}] $\lambda$6583 while fitting our combined LRS2 spectrum with the same three Gaussians plus a Lorentzian (Fig.~\ref{fwhmfit} lower panel). In so doing, we get a lower amplitude Gaussian for the broadened \Ha\ feature and a higher Lorentzian component than \citet{mauerhan18}. The FWHM of the Gaussian component turns out to be $\sim$$195 \pm 4$ \kms, while for the Lorentzian it is $\sim$$290 \pm 4.5$ \kms. Note that if we apply FWHM $\sim$130 \kms for the Gaussian component and fix its amplitude to be equal to that of the Lorentzian component, we recover the $\sim$400 \kms FWHM of the Lorentzian by \citet{mauerhan18}.

If we consider the base of the \Ha\ line profile instead of the FWHM, the width becomes $\pm 1000$ \kms. The total width at base is thus $\sim$2000 \kms. The FWHM is very sensitive to the amplitude of the Lorentzian, which depends on the spectral resolution and the amplitude ratio of the narrow and intermediate model components. It may be that the decelerated shock velocity is $\sim$1000--2000 \kms rather than $\sim$300--400 \kms. Note that the \Ha\ velocities we consider here could correspond to shock velocities, to intrinsic wind speeds, to broadening by electron scattering, or to the bulk motion of a cold, dense shell. We return to these issues in \S\ref{halphamod} and \S\ref{windbubble}

From our \dave{2017} HET spectra we measure $\ee{1.92 \pm 0.23}{-14}\ \ergcms$ for the observed total \Ha\ flux above continuum and $\sim 6.7 \pm 0.1$ for the \Ha/H$\beta$ flux ratio. Adopting $D = 22.8$ Mpc and $A_R = 0.342$ mag for the distance and $R$-band extinction \citep{Vinko17}, respectively, the total flux corresponds to an \Ha\ luminosity of $L$(\Ha)$=1.63 \pm 0.20 ~\times 10^{39}$ erg~s$^{-1}$ or $\log L = 39.21 \pm 0.05$. This luminosity agrees well with the one given in \citet{Vinko17} within the uncertainties, and does not confirm the recent brightening (by a factor of 3) in the \Ha\ luminosity of SN~2004dk found by \citet{mauerhan18}, even though our adopted distance of 22.8 Mpc is $\sim $10 \% larger than the one used by \citet{mauerhan18}. \dave{The flux measured from our 2019 spectrum is nearly identical to that of the 2017 spectrum.} The \Ha\ light curve is shown in Figure \ref{XradHalpha}.

\subsection{Analysis of [NII] $\lambda 6583$ line}
\label{profile}

\begin{figure}%[!ht]
\includegraphics[width=\columnwidth]{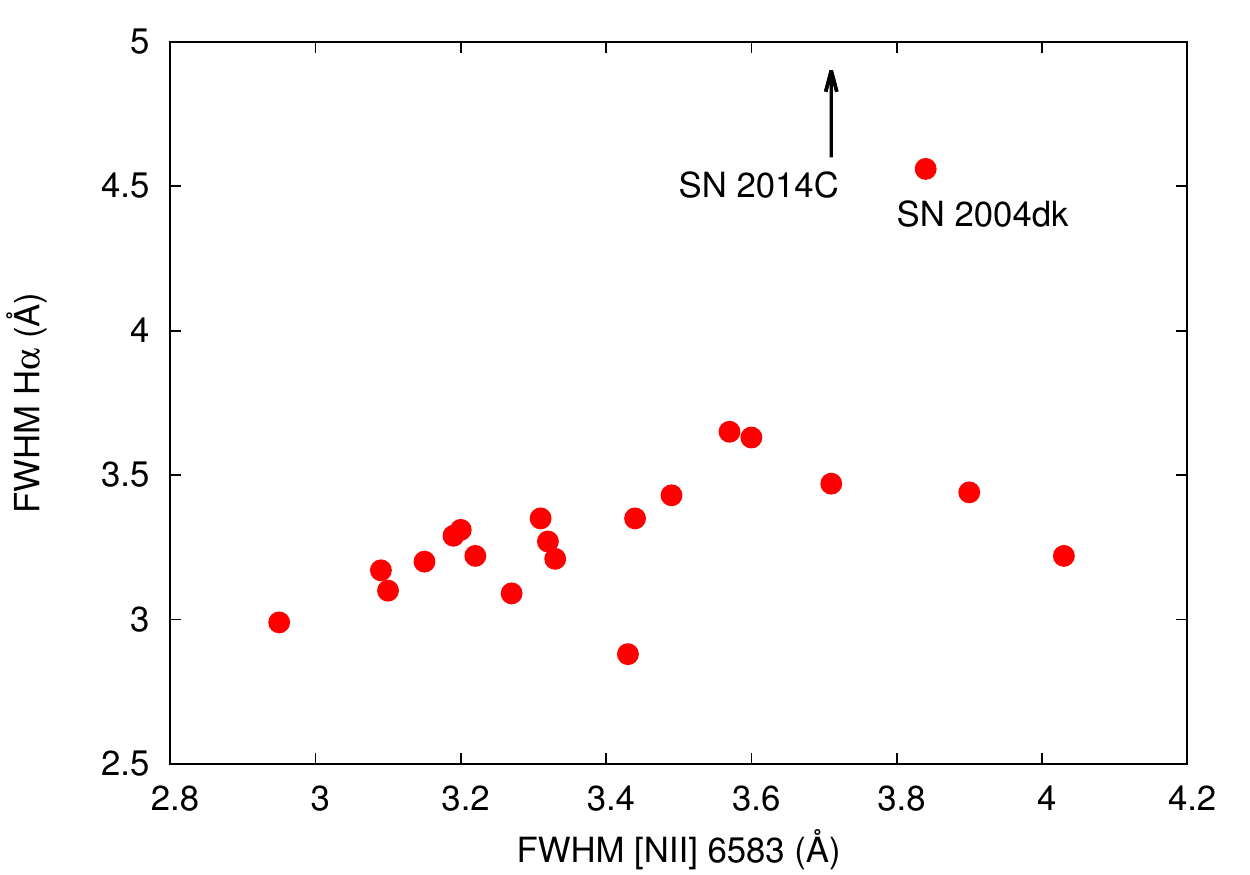}
\caption{A plot of the FWHM of H$\alpha$ compared to that of the [NII] $\lambda 6583$ line 
for a number of events in our sample. SN~2004dk and SN~2014C are distinctly more
broadened, with SN~2014C falling off the plot, as indicated by the arrow.}
\label{fwhm}
\end{figure}

Figure \ref{fwhm} gives a plot of the FWHM of H$\alpha$ compared to the FWHM of the [NII] $\lambda 6583$ line for a number of events in our sample for which spectra have been obtained (see Table~\ref{tab:fwhm} in the Appendix). Most of the events show FWHM of both lines in the range 3--4 \AA. Both SN~2004dk and SN~2014C (which falls off the top of the plot), are substantially more broadened than any of the others. 

\begin{figure}%[!ht]
\begin{center}
\includegraphics[width=3.5in]{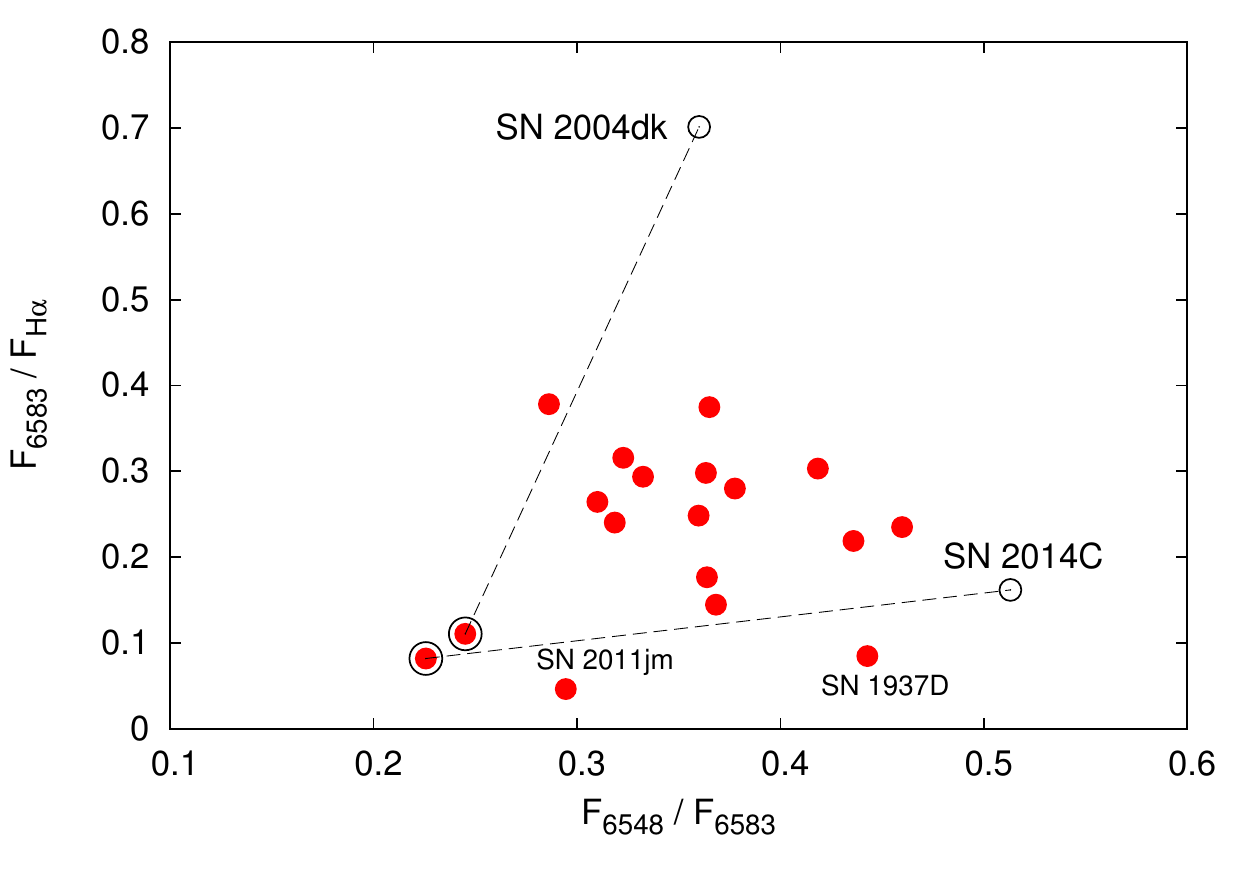}
%\includegraphics[width=3.5in]{diafi_lrs2_fluxr.png}
      %\vspace{-0.5cm}
      \caption{{\scriptsize
The ratio of the flux of [NII] $\lambda 6583$ to that of H$\alpha$ is plotted as a function of the ratio of the fluxes of the two [NII] lines, $\lambda 6548$ and $\lambda 6583$ for a number of events in our sample. For SN~2004dk and  SN~2014C, the open symbols represent data prior to late-time interaction and the solid points to data taken after the interaction had begun.
}}\label{ratio}
 \end{center}
\end{figure}

Figure \ref{ratio} gives the ratio of the flux of [NII] $\lambda 6583$ to that of H$\alpha$ as a function of the ratio of the fluxes of the two [NII] lines, $\lambda 6548$ and $\lambda 6583$, for a number of events in our sample. SN~2004dk and SN~2014C are shown by both open symbols based on data obtained prior to the onset of  late-time interaction and the solid symbols that represent the conditions after the onset of interaction. Note that the post-interaction \Ha\ flux contains both  the broad component formed in the ejecta behind the shock front as well as the narrow  component that can originate both in the CSM in front of the forward shock and in the  ISM of the host galaxy along the line of sight. The presence of the broad component  is the primary cause of the strong decrease of the [NII]/H$\alpha$ flux ratio during  the interacting phase. This is slightly different from the finding of \citet{Milisavljevic15} who defined the flux ratio based on only the narrow H$\alpha$ component. They found an {\it increase} of their flux ratio in SN~2014C during the interacting phase. Our definition, however, may be more suitable for objects like SN~2004dk in which the H$\alpha$ profile cannot be easily separated into broad and narrow components.

Figures \ref{fwhm} and \ref{ratio} suggest that both the FWHM comparison as well as the [NII]/H$\alpha$ flux ratio may serve as a diagnostic tool for identifying CSM-interaction. For example, in Figure \ref{ratio} the corresponding points for SN~1937D and SN~2011jm, the [NII]/H$\alpha$ flux ratio of which are below 0.1, are also highlighted. Although they have strong, but still narrow H$\alpha$ features, they are also candidates for CSM-interaction. 

\subsection{\Ha/H$\beta$ Ratio}
\label{HaHb}

The \Ha/H$\beta$ ratio is usually higher in interacting SNe than in \ion{H}{2} regions. Under the conditions of Case B recombination, which is thought to be approximately valid for most \ion{H}{2} clouds, \Ha/H$\beta$ $\sim$ 3.0 -- 3.3 is expected \citep{1974agn..book.....O}, while in interacting SN flux ratios 
$\sim$ 6 -- 10 are often observed. 

\begin{figure}%[!ht]
\centering
\includegraphics[width=\columnwidth]{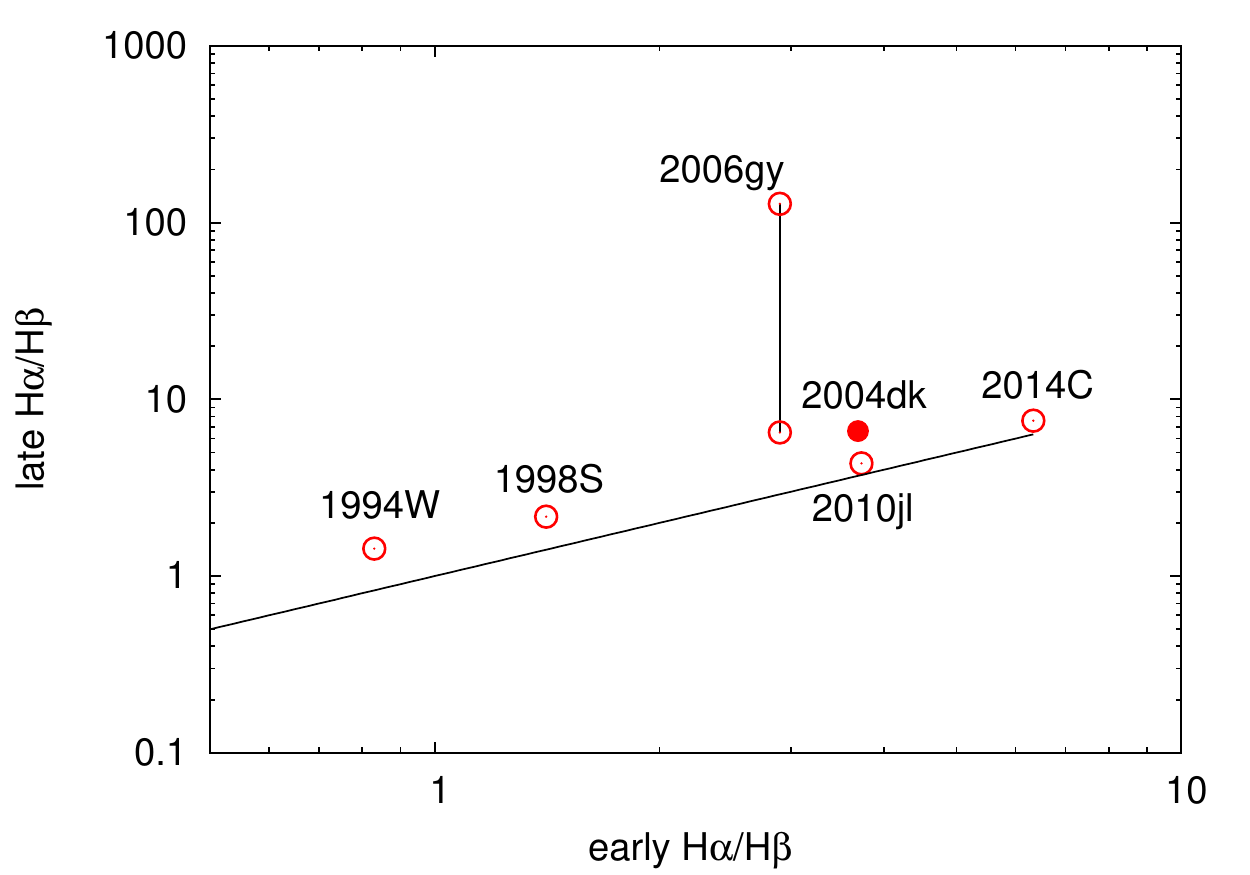}
\caption{The ratio of the flux of \Ha\ to that of H$\beta$ in the late, strong-interaction phase is plotted as a function of the ratio in early phases for several interacting SNe. The sloping black line represents a ratio of unity. SLSN~2006gy has two values connected with a vertical line. The lower point corresponds to the interaction phase around maximum light; the upper point corresponds to a later phase when the interaction is even stronger. SN~2004dk (filled circle) shows increased \Ha/H$\beta$ $\sim 6$ measured from our late-phase spectrum (Figure \ref{full_04dk}) compared to the earlier value of $\sim3.7$ as seen from the 2005 Keck spectrum (Figure~\ref{hal_04dk}).}
\label{fig:HaHb}
\end{figure}

Figure \ref{fig:HaHb} shows the \Ha/H$\beta$ flux ratios of several interacting SN including SN~2004dk. The flux ratio before the CSM-interaction (or at the very beginning of the interaction in case of SN that are discovered as Type IIn) is plotted on the horizontal axis, while the ratio during the strong interaction phase is shown on the vertical axis. Note that the superluminous SN~2006gy has two values, the lower point corresponding to the interaction phase around maximum light, while the other point corresponds to the late-phase, even stronger interaction. The sloping black line shows the 1:1 relation. It is seen that after the shock wave penetrates deep into the CSM, the \Ha/H$\beta$ ratio always become higher than before. SN~2004dk (plotted with a filled circle) is no exception: it also shows increased \Ha/H$\beta$ $\sim 6$ measured from our late-phase spectrum (Figure \ref{full_04dk}) compared to the earlier value of $\sim3.7$ as seen from the 2005 Keck spectrum (Figure~\ref{hal_04dk}). Note that there seems to be an apparent correlation between the hydrogen flux ratios before and after the CSM-interaction: SN showing larger \Ha/H$\beta$ in early phases tend to have higher flux ratio later. The very low number of events prevents a more definite conclusion. This could be an interesting subject for further studies.
 
\subsection{X-ray Properties}
Since shortly after its discovery, SN 2004dk has been observed by several X-ray satellites.  A log of the observations is given in Table~\ref{tab:xrayobs}.

\begin{deluxetable}{lllrr}[htb!]
\tablecaption{X-ray Observations of SN 2004dk \label{tab:xrayobs}}
\tablehead{
\colhead{Satellite} & \colhead{ObsID} & \colhead{Date (UT)} & \colhead{Days After } & \colhead{Exp.} \\ 
\colhead{} & \colhead{} &  \colhead{} & \colhead{Discovery\tablenotemark{a}} & \colhead{(ks)}
}
\startdata
\xmm   & 0164560801  & 2004 Aug 12.5 & 11.3   & 13.7\tablenotemark{b} \\
\swift & 00035228001 & 2006 Jan 21.7 & 538.6  & 3.3 \\
\swift & 00035228002 & 2006 Jan 24.0 & 540.8  & 6.0 \\ 
\swift & 00035228003 & 2006 Jan 25.1 & 541.9  & 4.3 \\
\chandra& 11226      & 2010 Jan 18.8 & 1996.6 & 8.0 \\
\swift & 00088663001 & 2018 Apr 17.9 & 5007.8 & 0.6 \\
\swift & 00088663002 & 2018 Jun 30.3 & 5081.1 & 3.7 \\
\chandra& 21349      & 2019 Jan 11.9 & 5276.8 & 9.4 \\
\enddata
\tablenotetext{a}{2004 Aug 01.19 \citep{2004CBET...75....1G}}
\tablenotetext{b}{for MOS2; see text for details.}
\end{deluxetable}

SN 2004dk was observed by the \xmmn\ as a Target of Opportunity requested by D.\ Pooley.  \xmm\ took the observation 11 days after discovery on 2004 Aug 12 for 17.6 ks of exposure \citep{Pooley07}.  We downloaded the ODF files from the \xmm\ Science Archive and reprocessed them using the latest calibration files (as of 2018 Jun).  We identified several large background flares in the data.  After filtering these out, there was 7.9 ks of exposure in the pn camera, 12.7 ks exposure in the MOS1 camera, and 13.7 ks exposure in the MOS2 camera.  We extracted source counts and spectra in the 0.4--8 keV band using a 15\arcsec\ source region centered on the position of SN 2004dk and a 1\arcmin\ source-free background region to the south.  We obtained net counts of $17.5\pm8.8$, $8.6\pm5.7$, and $27.6\pm7.2$ counts in the pn, MOS1, and MOS2 cameras, respectively, for a combined net counts of $53.7\pm12.7$.  

We fit all three source and background spectra simultaneously using the modified \citet{1979ApJ...228..939C} statistic cstat in Sherpa \citep{2006SPIE.6270E..1VF}.  We first use an absorbed mekal model with absorption fixed at the value through the Milky Way in the direction of SN 2004dk of \ee{8.3}{20} \pcmsq, calculated from the  Effelsberg-Bonn HI Survey \citep{2016A&A...585A..41W} using the online tool  at the Argelander-Institut f\"{u}r Astronomie\footnote{https://www.astro.uni-bonn.de/hisurvey/AllSky\_profiles/}.  The best fit model, with a reduced cstat of 0.704, prefers a temperature at the upper limit of the model, $kT = 80$ keV.  Adding an additional absorption component neither improves the fit nor lowers the best-fit temperature.  An absorbed power-law model has a best-fit power-law index of $1.1\pm0.3$ for an identical reduced cstat of 0.704.  Such a hard spectrum might indicate a non-thermal emission process, as expected for a Type Ib SN at early times \citep{2006ApJ...651..381C}. The unabsorbed 0.4--8 keV fluxes of the two types of best-fit models are the same within errors (Table~\ref{tab:fx}) around \ee{2}{-14} \ergcms, which corresponds to a luminosity of $\Lx \approx \ee{1.4}{39}$ \ergsec.

The \swift\ satellite observed SN 2004dk three times in 2006 Jan (around day 540) for a total XRT exposure of 13.6 ks.  Using a 10\arcsec\ source region and 1\farcm25 background region, we find 4 total counts in the source region, with an estimated 0.55 background counts in the region in the 0.4--8 keV band.  Using the Bayesian method of \citet{1991ApJ...374..344K}, we calculate a 68.3\% confidence interval of [1.7, 5.8] net counts and a 99.7\% confidence interval of [0, 12.8] net counts.  Because the 3$\sigma$-equivalent confidence interval includes zero, we regard this as a marginal detection at best and take the flux of the best-fit model as an upper limit.  To obtain that flux, we extracted and simultaneously fit source and background spectra of the merged data using methods similar to those above.  The best-fit mekal temperature is $kT = 25$ keV with no meaningful constraints because of the low quality of the data, and the unabsorbed 0.4--8 keV flux from this model is \ee{2.1}{-14}\ \ergcms.

The \cxo\ observed SN 2004dk on 2010 Jan 18 (around day 2000) for 8.0 ks with the telescope aimpoint on the Advanced CCD Imaging Spectrometer (ACIS) S3 chip as part of a program to investigate strong, late-time CSM interaction in Type Ib/c SNe, presumably due do the SN shock catching up to the previously cast-off hydrogen envelope (PI: Pooley).  Data reduction was performed with the chandra\_repro script, part of the Chandra Interactive Analysis of Observations (CIAO) software.  We used CIAO version 4.9 and calibration database (CALDB) version 4.7.8.  Using a 1\arcsec\ source region and 1\arcmin\ background region, we detect 7 total counts in the source region, with an estimated 0.06 background counts in the region in the 0.4--8 keV band.  We extract and simultaneously fit the source and background spectra using similar methods as above.  As with the earlier \xmm\ data, these data prefer a best-fit mekal temperature at the limit of the model ($kT = 80$ keV), which yields a reduced cstat of 0.489, but the temperature has a large uncertainty with a 68\% confidence interval for $kT$ of [4 keV, 80 keV].  An absorbed power-law model has a best-fit power-law index of $1.3\pm0.6$ and has a reduced cstat of 0.489, also.  The unabsorbed fluxes of the best-fit models are the same within errors (Table~\ref{tab:fx}) around \ee{1.5}{-14} \ergcms, which corresponds to a luminosity of $\Lx \approx \ee{9.5}{38}$ \ergsec.

\swift\ observed SN 2004dk again on 2018 Apr 17, but the XRT exposure is too short (687 s) to detect the SN in X-rays or to set an interesting limit, with only 1 count in the 10\arcsec\ source region and an estimated background contribution of 0.02 counts.  On the basis of our \Ha\ detections, our team requested a longer \swift\ observation, which occurred on 2018 Jun 30 and has an XRT exposure of 3.7 ks.  Merging the observations, we find 5 total counts in the source region, with an estimated 0.16 background counts in that region in the 0.4--8 keV band.  Again using the Bayesian method of \citet{1991ApJ...374..344K}, we calculate a 68.3\% confidence interval of [2.9, 7.5] net counts and a 99.7\% confidence interval of [0.7, 14.9] net counts, indicating a secure detection.  We extract and simultaneously fit the source and background spectra and obtain a best-fit mekal temperature of $kT = 0.9\err{0.5}{0.2}$ keV or a best-fit power-law photon index of $2.5\pm0.9$, which is markedly softer than previous X-ray observations.  The unabsorbed fluxes of the best-fit models agree within errors (Table~\ref{tab:fx}) around \ee{5}{-14} \ergcms, which corresponds to a luminosity of $\Lx \approx \ee{3}{39}$ \ergsec.

\dave{\chandra\ observed SN 2004dk again on 2019 Jan 11 for 9.4 ks as part of a program to follow up on interesting \Ha-emitting SNe from our DIAFI and HET programs (PI: Pooley).  The data reduction, fitting procedures, source region, and background region are identical to the above.  In the 0.4--8 keV band, we detect 58 total counts in the source region, with an estimated 0.06 background counts.  The best-fit mekal temperature is $kT = 4.2\err{1.7}{1.0}$ keV.  A power-law fit gives a photon index of $1.7\pm0.2$.  The unabsorbed fluxes of the best-fit models are the same within errors (Table~\ref{tab:fx}) around \ee{1.2}{-13} \ergcms, which corresponds to a luminosity of $\Lx \approx \ee{7}{39}$ \ergsec.  SN 2004dk brightened by a factor of $\sim$8 between the \chandra\ observation in 2010 and the \chandra\ observation in 2019.}

\begin{figure}%[!ht]
\centering
\includegraphics[width=\columnwidth]{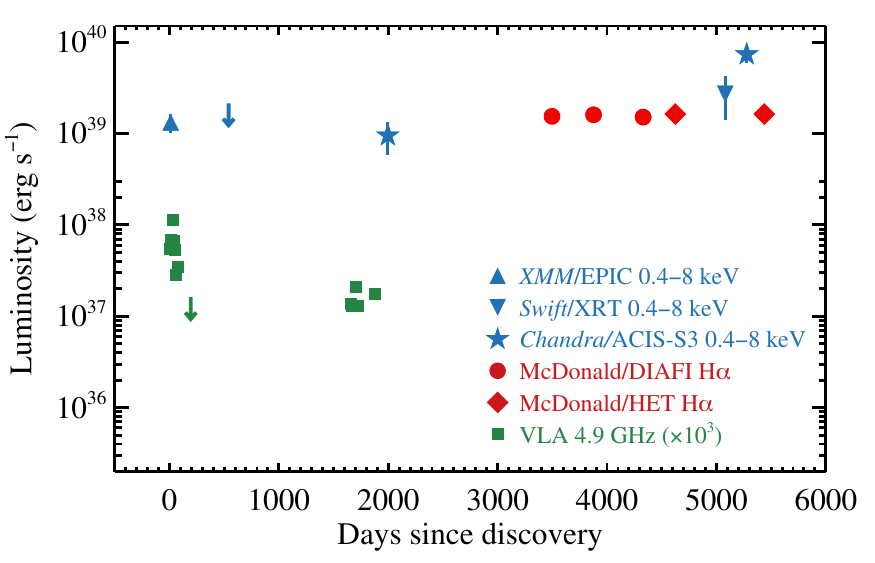}
\caption{Multiwavelength light curve of SN 2004dk.  X-ray measurements are shown in blue with symbols indicating the satellite (the upper limit is from \swift).  \Ha\ measurements are shown in red; circles indicate results from our narrowband imaging \citep{Vinko17}, and the diamond is from our HET spectra reported here.  Green squares are based on 4.9 GHz radio flux densities from the VLA reported by \citet{Stockdale09} and \citet{Wellons12} and converted to luminosities assuming a flat spectrum and 100 MHz bandpass.  }
\label{XradHalpha}
\end{figure}

The X-ray light curve is plotted with our \Ha\ light curve and radio data reported in \citet[][see below]{Wellons12} in Figure~\ref{XradHalpha}.

\begin{deluxetable}{lllrr}[htb!]
\tablecaption{X-ray Spectral Modeling of SN 2004dk \label{tab:fx}}
\tablehead{
\colhead{Observation} & \colhead{Model} & \colhead{PL Ind.} & \colhead{cstat / d.o.f.} & \colhead{\Fx\ [0.4-8\ keV]}  \\ 
\colhead{} & \colhead{} &  \colhead{or $kT$} & \colhead{} & \colhead{\ergcms}
}
\startdata
\xmm\ (2004) & mekal & 80\tablenotemark{a} & 3284.5 / 4664 & \ee{2.1\err{0.5}{0.5}}{-14}\\
                  & power law & 1.1\err{0.3}{0.3} & 3284.4 / 4664 & \ee{2.4\err{0.8}{0.5}}{-14}\\ \hline
\swift\ (2006) & mekal & 25\err{55}{22} & 182.7 / 1518 & $<\ee{3.4}{-14}$\tablenotemark{b}\\\hline
\chandra\ (2010) & mekal & 80\tablenotemark{a} & 506.7 / 1036 & \ee{1.5\err{0.6}{0.6}}{-14}\\
                  & power law & 1.3\err{0.6}{0.6} & 506.6 / 1036 & \ee{1.5\err{0.9}{0.5}}{-14}\\ \hline
 \swift\ (2018) & mekal & 0.9\err{0.5}{0.2} & 304.9 / 1518 & \ee{4.3\err{2.4}{2.1}}{-14}\\
                   & power law & 2.5\err{0.9}{0.9} &307.5 / 2528 & \ee{6.4\err{3.7}{2.6}}{-14}\\ \hline
\chandra\ (2019) & mekal & 4.2\err{1.7}{1.0} & 649.3 / 1036 & \ee{1.2\err{0.2}{0.2}}{-13}\\
                 & power law & 1.7\err{0.2}{0.2} & 655.4 / 1036 & \ee{1.3\err{0.2}{0.2}}{-13}
\enddata
\tablenotetext{a}{model maximum value}
\tablenotetext{b}{upper limit}
\end{deluxetable}

\section{Discussion}
\label{discuss}
 
\subsection{Early CSM Interaction}

\citet{Wellons12} observed the radio flux from SN~2004dk from about 8 days to about 80 days after explosion and fit its characteristics with a self-absorbed synchroton model. They found that the radio emission was consistent with the propagation of the SN shock into a steady-state wind with a density profile scaling as $\rho \propto r^{-2}$. They concluded that the electron density at the onset of the interaction was $n_e \sim 2.2\times10^4$ e$^-$ cm$^{-3}$ at a radius $\sim 5\times 10^{15}$ cm and that the shock propagated at a mean velocity of $0.2 c = 6\times10^4$ \kms\ over the first 80 days. For a wind velocity of 1000 \kms\ characteristic of a stripped core, they deduced a mass loss rate of $\dot{M} \approx 6\times 10^{-6}$ \msun\ yr$^{-1}$. They also determined an upper limit to the radio flux about 200 days after explosion that was consistent with the expected decline in a steady-state wind with the properties they had deduced that would have been less than the observed limits by about a factor of 3. 

Our \xmm\ observation around day 11 reveals a hard spectrum, indicating that thermal emission is probably not the source of the X-ray photons.  Rather, as suggested by \citet{2006ApJ...651..381C},  synchrotron and inverse Compton processes are likely responsible for the early X-ray and radio emission in Type Ib and Ic SNe.  A model X-ray light curve, presented in Figure 1 of \citet{2006ApJ...651..381C}, gives a predicted \Lx\ of $\sim$\ee{5}{38} \ergsec\ at an age of $\sim$11 days and assumes an optical light curve like that of SN~1994I.  Comparing the absolute $V$-band light curves of SN~1994I and SN~2004dk given in the Gold Sample of SN~Ib/c by \citet[][their Figure 7]{2011ApJ...741...97D}, SN~2004dk peaks about 1 magnitude more luminous than SN~1994I.  Our measurement of $\Lx \approx \ee{1.4}{39}$ \ergsec\ at $t=11$ days is $\sim$3 times higher than the model of \citet{2006ApJ...651..381C}.  Since the optical photons are inverse Compton upscattered to X-ray energies at early times, the X-ray flux should scale with the optical.  Since both our \Lx\ measurement and the peak optical luminosity of 2004dk are a factor of $\sim$3 higher than 1994I, the early X-ray emission detected by \xmm\ is likely explained by the synchrotron $+$ inverse Compton model of \citet{2006ApJ...651..381C}. This conclusion is consistent with the non-thermal origin of the contemporaneous radio emission. 

As a consistency check, we can estimate the bolometric luminosity of the shock at this time. Assuming spherical symmetry, the bolometric luminosity of the forward shock would be 
\begin{equation}
L_\mathrm{bol} = \frac{1}{2} \frac{\dot{M}}{v_w} v_\mathrm{sh}^3.
\label{lbol}
\end{equation}
Assuming $\dot{M} = 6\times 10^{-6}$ \msun\ yr$^{-1}$, $v_w = 1000$ \kms, and the shock to propagate at $6\times10^4$ \kms\ gives $L_\mathrm{bol} = 4.1 \times 10^{41}$ erg s$^{-1}$.
This estimate of $L_\mathrm{bol}$ is comfortably in excess of the observed X-ray luminosity at this epoch.
As a further check, we can estimate the X-ray luminosity if the X-rays were produced by a thermal mechanism, $\Lx = n_e^2 \Lambda V$, where $\Lambda$ is the cooling function and $V$ the radiating volume. We find $\Lx$ estimated in this way to be of order $10^{33}$ erg s$^{-1}$, far below the observed X-ray luminosity, giving further support to the non-thermal model for the early X-ray emission seen by \xmm\ at 11 days.  The later emission seen by \chandra\ at 2000 days is $\sim 10^2$ higher than predicted by the model of \citet{2006ApJ...651..381C}, indicating that another emission process is at work at late times.

\subsection{Late-time ``Rebrightening''}
\label{rebright}
\citet{Stockdale09} and \citet{Wellons12} detected radio emission from SN~2004dk 1660 days or 4.5 years after the explosion with a flux that was comparable to the 200-day upper limit. \citet{Wellons12} termed this later detection a ``rebrightening'' since it was about 40 times the flux expected from extrapolation of the steady-state wind model that fit the earlier detections. The radio rebrightening was attributed to the collision of the forward shock with denser circumstellar material, and our \chandra\ measurement of a high \Lx\ around day 2000 supports this interpretation.  
The beginning of the interaction of the ejecta with denser CSM material presumably happened some time before the radio rebrightening was detected. 

The CSM detected early likely arose from the fast wind from a stripped core, as discussed above. One hypothesis is that the rebrightening happened when the shock impacted the hydrogen-rich matter we detect; the radio rebrightening could represent mass that had been lost at a much earlier phase that was hydrogen rich and presumably ejected by a different physical process that was not necessarily a steady-state wind.  \citet{mauerhan18} proposed a model in which a fast wind blows a low-density bubble interior to a denser shell ejected at an earlier phase. The outer shell could be the material from a slower wind compressed by the later, faster wind \citep{Dwarkadas05,Dwarkadas10, Dwarkadas11} or material shed in a different process, for instance by common envelope evolution or a burst of episodic mass loss. 

Assuming that the epoch of the radio rebrightening represents an upper limit to the time when the shock began to interact with denser, circumstellar material, the boundary of the outer denser material associated with the rebrightening would be at $r < r_\mathrm{rb} \sim  1.4\times10^{17} v_\mathrm{sh,4}$ cm, where $r_\mathrm{rb}$ is the radius of the shock at the time of rebrightening and $v_\mathrm{sh,4}$ is the shock velocity in units of $10^4$ \kms. Adopting the shock velocity deduced by \citet{Wellons12}, $v_\mathrm{sh,4} = 6$ yields $r_\mathrm{rb} \sim 8.4\times10^{17}$ cm. For a self-similar shock for which $v \propto t^{(s-3)/(n-s)}$  where $n = 10$ is the density slope in the SN ejecta (estimates range from 7 to 12) and $s = 2$ is the density slope of the CSM, the velocity will scale as $v_\mathrm{sh,4} \propto t^{-1/8}$. 
For this mild deceleration, the shock would decelerate from $v_\mathrm{sh,4} = 6$ at 80 days to $v_\mathrm{sh,4} = 4$ at 1600 days.

This boundary to denser material must be greater than $5\times10^{15}$ cm, given the early radio observations of \citet{Wellons12}. A larger lower limit is given by the deduction of \citet{Wellons12} that the average velocity over the first 80 days is 0.2c. This gives a lower limit of $4\times10^{16}$ cm.

If the density continues to fall as $\rho \propto r^{-2}$ beyond the region of rebrightening, the bolometric luminosity from Equation \ref{lbol} would decline as $t^{-3/8}$. 
At 2000 days, the bolometric luminosity would be reduced by a factor of about 0.14 compared to that at 11 days and would still be comfortably more than the X-ray luminosity measured by \cxo\ at that later time, $9.5\times10^{38}$ \ergsec. If the density fell off less steeply, the bolometric luminosity would be even higher. We will revisit these issues in \S \ref{Xray} and \S \ref{windbubble}.

SN~2004dk was 3500 days or 9.6 years old when we first detected it as a bright \Ha\ point source on 2014 March 27 \citep{Vinko17}. The rebrightening was 1700 days earlier, and the collision with material denser than the inner fast wind must have happened earlier than that. The shock had at least 5.1 years to decelerate in the relatively denser material after the epoch of radio rebrightening. 
During this interval, the shock would have propagated with a mean velocity between the velocity we deduce at rebrightening, $v_\mathrm{sh,4} = 4$, and the velocity we deduce from 
the \Ha\ line width, $v_\mathrm{sh,4} = 0.03$. This would put the location of the shock at the epoch of the first broadened \Ha\ detection at $4.4\times10^{15}$ to $5.9\times10^{17}$ cm beyond the location of the rebrightening. The latter is almost surely an upper limit, so the implication is that the shock had not propagated substantially in radius from the time of the rebrightening.

Using a model of synchrotron self absorption to analyze the early radio data of SN~2004dk, \citet{Wellons12} concluded that the electron density at the early epoch was $n_{e,0} \sim 2.2\times10^4$ e$^-$ cm$^{-3}$ at a distance of $r_\mathrm{w,0} \sim 5\times10^{15}$ cm from the progenitor. They determined that the radio flux at the time of the rebrightening increased by a factor of $\sim 40$ compared to an extrapolation of the early radio data based on a steady-state wind. Assuming the magnetic field and the shock Lorentz factor, and hence a characteristic synchrotron frequency, do not change (assumptions that might be questioned) and equipartition of the magnetic field in the shock, \citet{Wellons12} argued that the synchrotron radio flux scales as $n_e^2 v_e^2$, where $v_e$ is some mean electron velocity. The increase in flux at the epoch of rebrightening thus suggested that the electron density of the CSM was higher by a factor of $40^{1/2} \sim 6.3$ compared to that obtained by extrapolating the earlier model to the location of the rebrightening. Assuming a constant mass loss rate at constant velocity, the density at the location of the rebrightening can be expressed as
\begin{equation}
\rho_\mathrm{rb} =  \frac{\dot{M}}{4 \pi v_w r_\mathrm{rb}^2}. 
    \label{rho_rebright}
\end{equation}
For $r_\mathrm{rb} = 1.4\times10^{17} v_{sh,4}$, this gives $\rho_\mathrm{rb} \sim 1.5\times10^{-23} v_{sh,4}^{-2}$ g cm$^{-3}$ or $\rho_\mathrm{rb} \sim 10^{-24}$ g cm$^{-3}$ for $v_{sh,4} = 4$. If the density jumps by a factor of 6.3, the density would then be $\rho_{sh} \sim 6\times10^{-24}$ g cm$^{-3}$.

\subsection{Constraints from the late-time \Ha\ Luminosity} 
\label{halphamod}

The \Ha\ luminosity provides constraints on the density of the material emitting that flux.
Taking the Case B recombination rate per unit volume to be $n_e n_p \alpha_B(T)$ with $n_e = n_p$ for pure hydrogen and the recombination coefficient for pure hydrogen to be \citep{1974agn..book.....O}
\begin{equation}
    \alpha_B(T) = 2.6\times10^{-13}~{\rm cm^3~s^{-1}}~T_4^{-0.8}
\end{equation}
where $T_4$ is the temperature in units of $10^4$K,
the recombination \Ha\ luminosity for a constant density cloud of radius $r$ is then
\begin{equation}
 L_{\Ha} =  n_p^2 \alpha_B(T) (4/3) \pi r^3 (h\nu_{\Ha}).
\end{equation}
With $h\nu_{\Ha} = 3\times10^{-12}$ erg, and $n_p = \rho N_0$, this can be expressed as
\begin{equation}
   L_{\Ha} = 1.2\times10^{24}~{\rm erg s^{-1}}~r^3 \rho^2 T_4^{-0.8}.  
\end{equation}
If the recombination flux comes from a thin shell, then the corresponding expression would be
\begin{equation}
   L_{\Ha} = 3.6\times10^{24}~{\rm erg s^{-1}}~r^3 \frac{\delta r}{r} \rho^2 T_4^{-0.8}.  
   \label{haflux}
\end{equation}
where $\delta r/r$ is the fractional thickness of the thin shell.

In \S\ref{profile} we found the \Ha\ flux for SN~2004dk to be $1.63\times 10^{39}$ erg s$^{-1}$ at 4626 days. \citet{mauerhan18} give the estimated \Ha\ luminosity for several related events as $2.5\times10^{39}$ erg s$^{-1}$ at 1327 days for SN~2014C, $1.3\times10^{39}$ erg s$^{-1}$ at 2218 days for PTF11iqb, and $2.2\times10^{38}$ erg s$^{-1}$ at 1883 days for SN~2009ip. Note that all these values are rather uncertain because of the uncertain distances and flux calibration issues. 

The location of the \Ha\ emission in each of these cases is uncertain. The propagation of the SN shock is uncertain since the propagation velocity depends on the possible complex CSM density distribution (see \S\ref{windbubble}). For a characteristic average shock speed of $10^4$ \kms\ propagating for several thousands of days, the typical scale would be a few times $10^{17}$ cm for these four objects. The characteristic distances would be $4.0\times10^{17}v_\mathrm{sh,4}$ for SN~2004dk, $1.2\times10^{17}v_\mathrm{sh,4}$ for SN~2014C, $1.9\times10^{17}v_\mathrm{sh,4}$ for PTF11iqb, and $1.6\times10^{17}v_\mathrm{sh,4}$ for SN~2009ig, where $v_{sh,4}$ is the shock velocity in units of $10^4$ \kms.  From equation \ref{haflux}, we can then write the density as
\begin{equation}
\rho_{-20} = 16.7~L_{\Ha,38}^{1/2} r_{17}^{-3/2} \left(\frac{\delta r}{r}\right)^{-1/2} T_4^{0.4}.
\label{density}
\end{equation}
For each of the SNe, the quantity $\rho_{-20} r_{17}^{3/2} \left(\frac{\delta r}{r}\right)^{1/2} T_4^{-0.4}$ is then 67 for SN~2004dk, 83 for SN~2014C, 60 for PTF11iqb, and 25 for SN~2009ig. Note that the \Ha\ luminosity must be coming from the outer dense shell. The density we derive for the region between $r_\mathrm{rb}$ and the outer shell, $\sim 10^{-23}$~g cm$^{-3}$, is far too low to generate the flux we detect.

The resulting CSM densities are rather insensitive to the temperature, but are rather sensitive to the location of the \Ha\ emission and to the potential thinness of any shell emitting \Ha. If the shell has a thickness only $\sim$ 0.01 the radius of the shell, the densities would all be larger by an order of magnitude. It is then difficult to compare the various events quantitatively because the radius and shell thickness could be quite different for the various events. Given the uncertainties in the values of the scaling parameters, the densities deduced in this manner are considerably higher than that deduced above for SN~2004dk based on the radio synchrotron emission, $\rho_{-20} \sim 5.5\times 10^{-3}$ just interior to the density jump represented by the rebrightening. The deduced densities are roughly comparable to the mean density of molecular clouds. 
The density we deduce for SN~2004dk, is, however, rather consistent with the density estimated from [\ion{N}{2}] emission in \S\ref{specprop}, $\rho_{-20} \sim 30 - 120$ for an ionized plasma.

Beginning at 3500 days, we observed SN~2004dk with narrowband filters for 830 days 
with no substantial change in \Ha\ flux \citep{Vinko17}. If the Lorentzian FWHM of $\sim 300$ \kms\ that we measure for \Ha\ corresponds to a shock speed, this implies that the conditions in the shell were nearly constant over a radial scale of $\Delta r \approx 2.2\times10^{15}$ cm. This length scale could be nearly an order of magnitude larger if the base width of the \Ha\ line were used, but the corresponding displacement would still be small compared to the radius, $\sim10^{17}$ cm. Given that the radius could not have changed substantially over this time scale and assuming that the shell thickness did not change substantially, then Equation \ref{haflux} suggests that the quantity $\rho^2 T_4^{-0.8}$ is nearly constant over this time interval. Even that is suspect, given the uncertainties in clumping, asymmetry, and other factors.This argument would not apply if the 300 \kms\ is the CDS velocity, not the shock velocity. \citet{mauerhan18} argued that the \Ha\ flux may have increased in just the last ~1000 days after \citet{Vinko17}; however, independent measurements presented in \S\ref{04dk} do not confirm such a change in the flux of \Ha.

\subsection{Constraints from late-time X-ray emission}
\label{Xray}

It is difficult to estimate the fraction of the bolometric luminosity 
that will be emitted in X-rays as it was to estimate the \Ha\ flux, as just illustrated. The X-ray luminosity will depend on the optical depth among other factors. %} 
For SN~IIn, the X-ray flux can be substantially more than the flux in \Ha\ in some cases, but substantially less in others \citep[][Chapter14]{BW17}. In the case of 2004dk, they are comparable \dave{for a while, but the X-rays have risen substantially by $\sim$5000 days, while the \Ha\ has remained constant} (Figure \ref{XradHalpha}). This clearly shows that the X-rays could not give rise to the \Ha\ flux. The \Ha\ cannot arise from a medium ionized by X-rays. %X-rays could come 
The standard picture of X-rays produced by a reverse shock in CSM interaction is of limited applicability here because we have the SN ejecta first interact with a lower density CSM and then a higher density CSM.  The \chandra\ observation of high \Lx\ at $\sim$2000 days confirms the onset of late-time interaction, as seen in the radio.  It does not provide additional constraints on the onset of the late-time interaction, but may provide some constraints on densities and velocities in the context of a correct physical model.

The X-ray flux from the \swift\ observation around 5100 days \dave{and \chandra\ observation around 5300 days} seems softer and hence may be due to thermal processes. If so, there is then a constraint on the shock velocity because if the velocity of the shock in the shell is too low, we would not expect any X-ray emission at all. One would need 400 \kms\ to get the temperature above $10^6$~K, and 500 \kms\ or more to get even very soft X-ray emission in the 0.2-0.2 keV band. The FWHM of the \Ha\ may be too low to produce thermal x-rays, although the full width of the base of the line (2000 \kms) would represent a sufficient velocity; this may indicate that the base of the \Ha\ line gives a better indication of the shock velocity than the FWHM. Another complication is that the shell we envisage might be clumpy. It could be that the shock propagates more rapidly in more rarified regions and slower in denser regions and that the thermal X-rays are not produced in the same region as the \Ha\ although both are produced in the dense shell. 
\dave{There are several ways to break spherical symmetry, but that is beyond the scope of the current paper; we will continue to acquire more  observations and address this issue in a future publication.}

\subsection{Wind-Bubble Model}
\label{windbubble}

The density in the vicinity of the rebrightening event derived in \S\ref{rebright}, $\rho \sim 5.5\times10^{-23}$ g cm$^{-3}$, is incommensurate with the density derived from the \Ha\ luminosity
in \S\ref{halphamod}, $\rho \sim 2.6\times10^{-19}$ g cm$^{-3}$ (modulo various scaling parameters). In addition, the estimated mass loss rate that could give the density jump estimated by \citet{Wellons12} in \S\ref{rebright}, $\dot{M}_\mathrm{o} \sim 3.8\times 10^{-7}$ \msun\ yr$^{-1}$, seemed unphysically low for plausible mass-loss processes that could produce a slow wind or other processes at the appropriate radius. These apparent discrepancies may be accommodated in the context of a model in which the SN explodes in a circumstellar medium that has been previously structured by interacting fast and slow winds, a wind-bubble model. 

Early numerical models of the interaction of a wind with the interstellar medium, including effects of conduction, were given by \citet{Weaver77}. \citet{CL89} and \citet{TT91} explored models of a SN exploding into such a ``prepared" CSM. \citet{Dwarkadas05}, \citet{Dwarkadas10}, and \citet{Dwarkadas11} investigated a SN exploding in a CSM in which a fast wind interacted with a previous slow wind which itself interacted with the surrounding ISM. This is a natural environment in which to consider our observations of SN~2004dk.

From the inside out, the structure of the CSM in a wind-bubble model has an inner fast wind that is plausibly due to a nearly steady-state wind from the stripped envelope, hydrogen-deficient progenitor. 
In this case, the inner CSM has the characteristic density profile $\rho_{i} \propto r^{-2}$. This inner wind structure ends at a termination shock where the freely expanding fast wind is shocked by the reverse shock resulting from the fast wind interacting with the outer, slower wind. This shock results in a density jump of about a factor of 4 for a strong shock in a perfect monatomic gas. Beyond the termination shock, the density slowly rises to the inner boundary of the contact discontinuity between the inner fast wind and the outer slow wind. 

Under common circumstances, the matter in the vicinity of the outer shock where the fast wind collides with the slow wind will be strongly radiative, leading to cooling and the formation of a cold, dense shell (CDS). At that point, the CSM density typically jumps by a factor of order $10^3$. If the CDS forms from the interaction of two steady state winds, the resulting velocity of the CDS is constant, $v_\mathrm{CDS} \propto (L_\mathrm{wind}/\rho)^{1/3}$ \citep{Weaver77,Dwarkadas05}. Beyond the CDS, the forward shock of the wind interaction propagates into the outer portions of the dense, slow wind \citep[][Figure 4]{Dwarkadas05}. 

Note, critically, that the density structure between the termination shock and the CDS corresponds to non-homologous structure and to no specific mass loss rate. The resulting overall density structure is the result of neither a steady state wind nor homologous expansion; it is a characteristic of interacting flows. 

Note also that if the fast inner wind is hydrogen deficient then so also is the region of the fast inner freely-expanding wind and the region between the termination shock and the CDS. In this picture, the \Ha\ emission we detect cannot be from near the termination shock, but must be from within or beyond the CDS, presuming the outer slow material to be of solar composition. 

This basic picture can be complicated by a number of factors, including Rayleigh-Taylor instabilities, as discussed by \citet{Dwarkadas07}, but captures the essence of the situation. The SN then explodes into the rather complex ``prepared" CSM.

When the SN explodes, the ejecta shock runs relatively quickly through the inner freely-expanding wind, somewhat more slowly through the nearly constant-density region between the termination shock and the CDS, and then very slowly through the CDS before breaking out into the outer, undisturbed slow wind. The passage of the SN shock substantially changes the density structure of the medium. When the ejecta shock hits the termination shock, a transmitted shock is sent forward into the nearly constant density portion of the wind bubble until it encounters the CDS. A new reflected shock propagates back into the ejecta, overcoming the original reverse shock and continuing into the ejecta. The original wind termination shock is obliterated. 

There is still a jump of a factor of several in density at the location of the new reflected shock. This jump is thus very reminiscent of the apparent jump in density at the epoch of the radio rebrightening 4.5 yr after the explosion of SN~2004dk. The increase in density is likely to be a factor of 4 representing the jump over a strong stock rather than the uncertain factor of 6.3 formally derived by \citet{Wellons12}. 

Given the basic structure of the CSM prepared by the interaction of fast and slow winds, we can outline the resulting characteristics of the matter in the wake of the subsequent SN shock. We normalize to the initial condition presented by \citet{Wellons12}: electron density $n_e \sim 2.2\times10^4$ e$^-$ cm$^{-3}$ at a radius $\sim 5\times 10^{15}$ cm and a mean velocity of $0.2 c = 6\times10^4$ \kms\ over the first 80 days. As noted in \S\ref{rebright}, the shock is likely to have slowed to a velocity of $v_\mathrm{sh,rb,4} = 4$ after 1600 days. We take a mean velocity during this interval to be $<v_\mathrm{sh,4}> = 5$. Adopting this shock velocity gives the location of the rebrightening as $r_\mathrm{rb} \sim 6.9\times10^{17}$ cm. 

From Equation \ref{rho_rebright}, the density at the outer limit of the fast wind at $r_\mathrm{rb}$ will be $\rho_\mathrm{rb-} = 6.3\times10^{-25}$~g cm$^{-3}$. Assuming the strong shock at the transition from the fast to the outer slow wind results in a density jump of a factor of 4, the density just beyond this jump at $r_\mathrm{rb}$ will be $\rho_\mathrm{rb+} = 2.5\times10^{-24}$~g cm$^{-3}$. 

Across the presumed density discontinuity, pressure, and hence $\rho v^2$, is nearly constant. The shock velocity just beyond the density jump will then decrease by a factor of 2 to  $v_\mathrm{sh,4} \sim 2$. For a fast wind of constant mass loss rate colliding with a slow wind of constant mass loss rate, the density profile between the transition shock at $r_\mathrm{rb}$ and the outer dense shell will be nearly constant. A similarity solution then gives $v_\mathrm{sh} \propto t^{-3/n} = t^{-0.3}$. The velocity of the shock will then decelerate in this medium as 
\begin{equation}
v_\mathrm{sh} = v_\mathrm{sh,rb}\left(\frac{t_{shell}}{t_{rb}}\right)^{-0.3} \approx  v_\mathrm{sh,rb,4}\left(\frac{3500}{1600}\right)^{-0.3}
= 0.8 v_\mathrm{sh,rb,4},
\label{v_rb_sh}
\end{equation}
where we have taken the epoch when the shock hits the shell, $t_{shell}$, to be the time of our first observation of bright \Ha. By the time the shock hits the boundary of the dense outer shell, it will thus have decelerated to a velocity of $v_\mathrm{shell,4} \sim 1.6$.

Likewise, a similarity solution in a constant density medium gives the location of the shock to be $r_\mathrm{sh} \propto t^{(n-3)/n} = t^{0.7}$. The location of the shock in the constant density medium when the shock arrives at the boundary of the outer dense shell is then approximately
\begin{equation}
r_\mathrm{shell} = r_\mathrm{rb}\left(\frac{t_{shell}}{t_{rb}}\right)^{0.7} \approx  r_\mathrm{rb}\left(\frac{3500}{1600}\right)^{0.7} = 1.7 r_\mathrm{rb}.
\label{r_rb_sh}
\end{equation}
With $r_\mathrm{rb} \sim 6.9\times10^{17}$ cm, the location of the inner edge of the shell is thus estimated to be at $r_\mathrm{shell} \sim 1.2\times10^{18}$~cm.

When the shock enters the outer dense shell, we can again invoke the behavior across a density discontinuity, $\rho v^2 \sim$ constant so that the velocity within the shell, $v_\mathrm{in}$, can be written
\begin{equation}
v_\mathrm{in} = v_\mathrm{shell}\left(\frac{\rho_\mathrm{rb}}{\rho_\mathrm{in}}\right)^{1/2}.
\label{v_in}
\end{equation}
We take the estimate of the density within the dense shell as constrained by the \Ha\ luminosity (Equation \ref{density}) to be
\begin{equation}
\rho_\mathrm{in} = 6.7\times10^{-19} r_{17}^{-3/2} \left(\frac{\delta r}{r}\right)^{-1/2} T_4^{0.4}
\label{rhoHalpha}
\end{equation}
and hence, with the density on the exterior of the density jump at the rebrightening of $\rho_{rb+} \sim 2.5\times10^{-24}$~g cm$^{-3}$, we find
\begin{equation}
v_\mathrm{in} \sim 200~{\mathrm km~s^{-1} } \left(\frac{\delta r}{r}\right)^{1/4} T_4^{-0.2}
\label{v_in}
\end{equation}
This result is roughly compatible with the FWHM that we measure for the \Ha\ line. Note that $\delta r/r$ must be considerably less than unity to comport with the thin shell approximation we have implicitly made in the analysis. 

The properties we derive for SN~2004dk are thus qualitatively consistent with a wind-bubble model in which the SN exploded into a CSM comprising the previous interaction of fast and slow winds.

\subsection{The Slow Wind}   
\label{slow}

If the outer denser CSM material had been ejected as a wind at about 10 \kms as might be characteristic of a red supergiant or a common envelope, and the shocked material lies at about the distance we estimate for the outer dense shell, $r_\mathrm{shell} \sim 1.2\times10^{18}$~cm, then this material would have been ejected about 48,000 years ago, perhaps during helium core burning. Other matter from this slow wind could have been ejected earlier and be at larger radii, but we have no constraints on that. Another component of the slow wind would have been ejected later, but then swept up by the even later, fast wind. When the fast wind turns on, it compresses the slow wind, forming the dense shell. As the fast wind continues, it sweeps up more slow wind material. The mass of the shell increases with time as more slow wind material is swept up, but the velocity of the shell is nearly constant. A similarity solution \citep[][appendix]{Dwarkadas10} gives
\begin{equation}
v_\mathrm{shell} \approx \left(3 \frac{\dot{M}_\mathrm{fast}}{\dot{M}_\mathrm{slow}}v_\mathrm{slow}v_\mathrm{fast}^2\right)^{1/3}.
\label{vshell}
\end{equation}
With $\dot{M}_\mathrm{fast} = 6\times10^{-6}$ \msun\ y$^{-1}$ and $v_\mathrm{fast} = 10^3$ \kms\ from \citet{Wellons12}, $\dot{M}_\mathrm{slow} = 10^{-6}$ \msun\ y$^{-1}$, a typical mass-loss rate for a red supergiant, and $v_\mathrm{slow} = 10$ \kms, we find $v_\mathrm{shell} \approx 270$ \kms\ and the time for the shell to reach its current distance of $1.2\times10^{18}$ cm to be 1400 years. 
The onset of the fast wind would thus correspond roughly to core carbon burning, long before oxygen or silicon burning, depending on the ZAMS mass of the progenitor. 

From momentum conservation, the CSM mass that can decelerate a mass of shocked material $M_\mathrm{sh}$ from $v_0$ to $v_1$ is
\begin{equation}
M_\mathrm{CSM} \sim M_\mathrm{sh} \times \frac{(v_0-v_1)}{v_1} \approx \frac{v_0}{v_1} 
\end{equation}
for $v_0 >> v_1$ \citep[e.g.][]{2006ApJ...641.1051C}. If $v_0 \sim 10^4$ \kms\ and $v_1 \sim 300$ \kms this gives a mass ratio of $M_\mathrm{CSM}/M_\mathrm{sh} \sim 32$.

We can deduce the amount of slow, dense wind swept up by the fast inner wind at the location of the dense shell, $M_\mathrm{swept,shell}$, as
\begin{equation}
    M_\mathrm{swept,shell} = \dot{M}_\mathrm{slow} \frac{r_\mathrm{shell}}{v_\mathrm{w,o}} \sim 0.04~\msun\ \dot{M}_\mathrm{slow, -6} 
\end{equation}
assuming $r_\mathrm{shell} = 1.2\times10^{18}$, $v_\mathrm{w,o} = 10$ \kms, and where we have scaled the presumed slow wind to $\dot{M}_\mathrm{slow} = 10^{-6}$~\msun~yr$^{-1}$. This is a plausible amount of mass to be swept up from a red supergiant wind, given uncertainties about the mechanism. The total mass ejected in a red supergiant wind or other process could be even larger but remains unshocked and undetected.

\section{Conclusions and Future Work}   
\label{concl}

We have presented late-time optical and X-ray observations and analysis of SN~2004dk demonstrating that the SN is now strongly interacting with mass loss shed from the progenitor star prior to core collapse.  The line of \Ha\ is broadened to FWHM of about 300 \kms. In addition, SN~2004dk shows a ratio of \Ha\ to H$\beta$ of about 6, far stronger than typical \ion{H}{2} regions and another indication of interaction. SN~2004dk also shows evidence for [\ion{O}{1}] and \ion{He}{1} that may be heated or excited by the interaction.

We have presented \xmm, \chandra, and \swift\ observations that show a remarkably constant luminosity that is close to that emitted by \Ha. We argue that the early X-rays at 11 days and those at 2000 days are likely to be non-thermal, but that the X-rays detected by \swift\ \dave{and \chandra\ after 5000} days are probably thermal. We have no ready explanation for why the thermal and non-thermal luminosities are so similar, but note that other long-lasting interactions have also produced nearly constant X-ray luminosity.  The detection of thermal X-rays later than 5000 days implies a shock velocity greater than several hundred \kms\ and likely more than $\sim$1000 \kms.

We present a wind bubble model in which the CSM is ``pre-prepared" by a fast wind interacting with a slow wind. This model allows a reasonable estimate of the velocity manifested by the broadened \Ha\ in terms of the early mass loss and the shock velocity suggested by the radio observations. We estimate that the slow wind material currently in the vicinity of the dense shell was blown about 48,000 years ago, perhaps during the red supergiant phase of core helium burning, and that the fast wind started about 1400 years ago as the star ejected the last of its hydrogen envelope, perhaps during carbon burning. In this interacting-wind model, much of the outer density profile into which the SN explodes corresponds to no steady state mass loss process.  The dense shell arises naturally in a model of interacting winds and does not require a special epoch of temporarily enhanced mass loss \citep{shiode14,fuller17,nance18}; it is just a fast, low-density Wolf-Rayet wind interacting with a red supergiant wind.

In numerical models, when the forward shock breaks out of the CDS, its velocity increases to in excess of 1000 \kms\ before decreasing over a long time scale as it continues to propagate into the undisturbed outer slow wind. The late-time thermal X-rays may be consistent with that velocity as is the  width of the \Ha\ line at its base. If the shock has propagated beyond the CDS, radiation from the forward shock could photoionize the matter in the undisturbed wind. Given that the X-ray and \Ha\ are of nearly equal luminosity and that the X-ray emission has risen with no corresponding rise in \Ha, this cannot be an explanation for the \Ha\ flux. The absence of strong narrow emission lines might suggest the forward shock has not yet broken out of the CDS for SN~2004dk,  but evidence for this may be hidden in unresolved lines. The shock and associated excitation structure may be more complex if the shock is propagating through a turbulent, clumpy medium. It may be that the relatively narrow \Ha\ is generated from dense clumps, and the flux in the base of the line and the thermal X-rays are produced in more rarefied material between the clumps. The X-rays could also be produced by the reflected shock from the CDS going into a somewhat lower density  region inside of the shell.

We do not know when SN~2004dk began producing \Ha\ luminosity. It showed no obvious variation in \Ha\ flux in our narrowband imaging, but it may be in a long-lived phase of interaction with slow evolution. We note that other events in our sample show the opposite effect; they show variable \Ha\ flux, yet no obvious sign of line broadening.

It is interesting to know how SN~2004dk compares and contrasts with other stripped-envelope core collapse events, especially SN~Ib, that have and have not displayed evidence of late-time interaction with hydrogen-rich material. \dave{A systematic comparison is the subject of a later paper, but a} comparison of the \Ha\ flux of few related objects --- SN~2004dk, SN~2014C, PTF11iqb, and SN~2009ig ---  at various phases ranging from 1300 to 4600 days is given in \S \ref{halphamod}. Characteristic shell radii are $\sim 10^{17}$ cm. The density of the region emitting \Ha\ is rather insensitive to the temperature, but is sensitive to the location and the assumed thickness of the shell.

The event that triggered this particular line of investigation for SNIb/c was SN~2001em. SN~2001em may have been discovered before maximum light \citep{2006ApJ...641.1051C}. Interaction was detected in the radio about 770 days after the estimated date of explosion, and X-rays were detected on day 937 \dave{at the level of $\Lx \approx 10^{-41}\, \ergsec$}. Hydrogen emission was observed on day 970. These time scales are commensurate with those of SN~2004dk, but, as for SN~2004dk, the interaction must have begun at an indeterminate time prior to these observations. \citep{2006ApJ...641.1051C} adopt a model in which a dense, massive, 2.7 \msun, circumsteller shell is \dave{formed by a period} of very high mass loss, 2--10$\times10^{-3}$ \msun\ y$^{-1}$, \dave{followed by a fast wind which swept up the lost mass into a dense shell and accelerated it}. They derive a rather modest final velocity of the shell prior to the explosion of $\sim$30--50 \kms. In the wind bubble model we invoke here, the velocity of the shell is essentially constant as given by Equation \ref{vshell} for which we find the shell velocity for SN~2004dk to be $270$ \kms. Given the similarity \dave{of the models used for each SN, it is likely that the differences in derived shell velocities and densities are related to the different X-ray luminosities of the two SNe and possibly to different timescales for the onset of the interaction with the shell.}

Of particular interest is the well-studied case of SN~2014C. Interaction was already conspicuous in optical spectra of SN~2014C on day 113 when it emerged from solar hiatus \citet{Milisavljevic15,Margutti17}, and the radio brightened in that event at about 200 days \citet{Anderson17}. The beginning of the interaction in SN~2004dk is uncertain. There is no sign of it in the \Ha\ spectrum of \citep{Shivvers17} at 284 days, but the evidence is robust in radio emission at about 1600 days. Thus, SN~2004dk might have turned on in roughly the same time frame, hundreds of days, as SN~2014C, but might have been delayed by a factor of 10 longer, as we have assumed in much of our analysis here. This difference in time of onset might be accommodated with appropriate choices of wind timing, mass loss rates and wind velocities without demanding a completely different model for the nature of the CSM in SN~2014C in the context of a similar interacting winds model. 

Early spectra of SN~Ib often show evidence for a small mass of hydrogen \citep{Parrent16,BW17}. Weak high-velocity components of \Ha\, H$\beta$ and H$\gamma$ were observed, with velocities of $-15,000$ \kms\ in the double-peaked SN~2005bf \citep{Maund07a}. SN2014C showed some evidence for an extended high-velocity \Ha\ absorption feature near maximum, suggesting that the progenitor star was not completely stripped of hydrogen \citet{Milisavljevic15}. There seems to be no such evidence for SN~2004dk. On the contrary, the He I emission lines of SN~2004dk appear to be stronger than those of SN~2014C.  SN~2014C and SN~2004dk might thus have had some difference in their progenitor evolution. The CSM between the progenitor and the CDS might have been more helium rich in SN~2004dk. SN~2014C developed broadened lines of 
[\ion{O}{3}] $\lambda\lambda$4959, 5007 282 days after the explosion. 
The implied velocity, in excess of 3500 \kms, implied that this oxygen was in the SN ejecta that had been heated by the reverse shock. There is no obvious evidence that these oxygen lines have multiple kinematic components. SN~2004dk has, as yet, not shown such high-velocity oxygen, implying some difference in the strength of the reverse shock or some means to obscure the inner regions of the ejecta.  

Although Type IIn SN~1996cr exploded at $\sim3.7$ Mpc, it was not discovered until several years past explosion \revision{\citep{2008ApJ...688.1210B}}. X-ray and radio data were not obtained early on, but in many cases were retrieved from archival data, because the AGN at the center of the host galaxy (ESO 97-G13) had often been observed. The radio flux showed a sharp increase starting around 700 d, a flattening of the light curve around 2000 d, and a decrease and turnover around 5000 d \citep{2013MNRAS.431.2453M}. The X-rays have fairly restrictive upper limits at 700 and 900 d, but show a definite slowly rising flux by 2000 days that ultimately turns over. The X-ray and radio behavior was addressed by \citet{Dwarkadas10} who adopted a wind-bubble model with a pre-SN CSM consisting of a freely-expanding steady-state wind ending in a strong wind-termination shock, beyond which is a constant-density shocked-wind region and then the CDS formed by the fast/slow wind interaction, as outlined in \S\ref{windbubble}. To match the X-ray light curve, \citet{Dwarkadas10} carried out numerical simulations of the SN ejecta colliding with the dense shell of material formed by the interaction of the fast and slow winds. They concluded that the CDS had a density of $1.28\times10^{-19}$ g cm$^{-3}$ extending from $1.0\times10^{17}$ to $1.5\times10^{17}$ cm implying a mass of 0.64 \msun. The wind-termination shock is located at $1.6\times10^{16}$ cm, and the shocked-wind region has a constant density of $8.2\times10^{-22}$ g cm$^{-3}$. 

In the models of \citet{Dwarkadas10}, the SN forward shock reaches the CDS in about 3 years and departs it at about 7 years past explosion. By the time it was observed in the optical, SN~1996cr displayed hydrogen emission and was classified as a Type IIn, but the modelling revealed that the early evolution was best explained with a wind bubble model with a fast wind just before explosion. The deduced features of SN~1996cr are thus similar to SN~2004dk. A wind-bubble model formed by interacting fast and slow winds gives a satisfactory, rather tightly constrained, agreement with the data, suggesting that SN~1996cr resembled a SN~Ib/c at birth. Once again, there is no need in this instance to assume a sudden impulsive mass ejection to account for the dense shell. 

By contrast, SN~1986J seems to have been a SN~IIn from the time of explosion. Its X-ray flux \revision{\citep[e.g.,][]{1998ApJ...493..431H}} and optical spectra \revision{\citep[e.g.,][]{2008ApJ...684.1170M}} are typical of \dave{an X-ray luminous} SN~IIn. In addition, it was observed from onset in the radio with VLBI \revision{\citep[][and references therein]{2002ApJ...581.1132B}} so we know that it is very unlikely that the progenitor was a Wolf-Rayet star.

The progenitor of an SN~Ib/c has lost its hydrogen and maybe its helium envelope. This does not make it specifically a Wolf-Rayet star \citep{BW17}, but certainly suggests it could blow a fast, low-density wind. If such a wind is produced in the very late phase of evolution, it will sweep up the earlier lower-velocity RSG wind to form a dense shell. At some point the SN shock must collide with this shell. The time at which that happens will be decided by the properties of the star that determine the wind properties at various stages.

For all objects where we detect \Ha\ emission at the site of a SN, we are gathering data at other wavelengths (radio, mid-infrared, and X-ray) and optical spectroscopy to construct a more detailed picture of the immediate environment of these objects. We are currently obtaining joint \chandra/VLA observations of another eight Type I SNe that have also shown signatures of late-time interaction discovered in our ground-based optical survey at McDonald Observatory.  We will also be continuing our ground-based efforts with the aim of doubling the sample of old Type I SNe observed in \Ha.

\acknowledgments

We thank Sarah Wellons and Christopher Stockdale for discussions of the radio emission of SN~2004dk and Nick Chugai for discussions of \Ha\ emission. We also thank the referee whose cogent reports substantially improved the paper. Support for this work was provided in part by the National Aeronautics and Space Administration through Chandra Award Numbers GO0-11007A and GO9-20065A issued by the Chandra X-ray Center, which is operated by the Smithsonian Astrophysical Observatory for and on behalf of the National Aeronautics Space Administration under contract NAS8-03060. This research was supported in part by NSF AST-1109801 and by NSF AST-1813825. JV is supported by the project ``Transient Astrophysical Objects" GINOP 2.3.2-15-2016-00033 of the National Research, Development and Innovation Office (NKFIH), Hungary, funded by the European Union. VVD’s research is supported by National Aeronautics and Space
Administration Astrophysics Data Analysis program grant NNX14AR63G (PI Dwarkadas) awarded to the University of Chicago.

\begin{appendix}

The measured fluxes and FWHM values of the \Ha\ and the [\ion{N}{2}] $\lambda \lambda$6548,6583 features for our sample of SNe are given in Table \ref{tab:fwhm}. All spectra were taken with HET/LRS2. Note that absolute fluxes are not calibrated; only the relative fluxes are reliable.

%\newpage

\begin{table}[h]
\begin{center}
\caption{Measured fluxes and FWHM values of the \Ha\ and
the [\ion{N}{2}] $\lambda \lambda$6548,6583 features for
SNe observed in our program.}
\label{tab:fwhm}
\begin{tabular}{lccccccc}
\hline
SN & Date & $F_\lambda$(\Ha) & $F_\lambda$(6548) & $F_\lambda$(6583) & FWHM(\Ha) & FWHM(6548) & FWHM(6583) \\
   &      & (cgs)\tablenotemark{a} & (cgs) & (cgs) & (\AA) & (\AA) & (\AA) \\
\hline
PTF11kx & 2016-12-07 & 7.4 & 0.8 & 1.7 & 3.44 & 4.13 & 3.90 \\
2000cr & 2017-04-23 & 17.1 & 2.3 & 6.4 & 3.43 & 3.96 & 3.49 \\
2000ew & 2017-03-28 & 10.9 & 0.8 & 3.0 & 3.10 & 3.10 & 3.10 \\
2004ao & 2017-03-27 & 0.48 & 0.06 & 0.14 & 2.88 & 3.19 & 3.43 \\
2004dk & 2017-03-31 & 22.1 & 0.9 & 2.3 & 4.56 & 5.94 & 3.84 \\
2004gn & 2016-12-26 & 5.6 & 0.6 & 1.6 & 3.21 & 3.40 & 3.33 \\
2005kl & 2016-12-10 & 119 & 11.5 & 35.5 & 3.65 & 3.67 & 3.57 \\
2007af & 2017-03-28 & 0.98 & -- & 0.16 & 3.22 &  --  & 4.03 \\
2007gr & 2017-09-12 & 7.0 & 0.6 & 1.9 & 2.99 & 3.24 & 2.95 \\
2008dv & 2016-09-09 & 7.9 & 1.0 & 2.9 & 3.35 & 3.25 & 3.31 \\
2011jm & 2017-02-25 & 13.5 & 0.21 & 0.63 & 3.17 & 3.43 & 3.09 \\
2012fh & 2016-12-23 & 0.78 & 0.06 & 0.19 & 3.31 & 2.69 & 3.20 \\
2012P  & 2017-03-31 & 25.9 & 2.6 & 8.2 & 3.22 & 3.17 & 3.22 \\
2014C\tablenotemark{b} & 2017-05-24 & 17.2 & 1.0 & 3.6 & 3.47 & 4.96 & 3.71 \\
2014C\tablenotemark{c} & 2017-05-24 & 19.2 & -- & -- & 50.51 &  -- &  -- \\
2014L  & 2017-04-22 & 66.8 & 3.4 & 9.6 & 3.63 & 4.08 & 3.60 \\
1937D  & 2016-08-04 & 11.5 & 0.36 & 0.90 & 3.09 & 3.60 & 3.27 \\
1979B  & 2017-03-27 & 1.52 & 0.08 & 0.23 & 3.29 & 2.84 & 3.19 \\
1981B  & 2017-01-08 & 1.29 & 0.07 & 0.22 & 3.20 &  --  & 3.15 \\
1983I  & 2017-01-26 & 20.0 & 1.67 & 5.10 & 3.35 & 3.61 & 3.44 \\
1985F  & 2016-12-06 & 21.6 & 1.21 & 3.53 & 3.27 & 3.16 & 3.32 \\
\hline 
\hline
\end{tabular}
\end{center}
\tablenotetext{a}{in $10^{-15}$ \ergcms\ units}
\tablenotetext{b}{Narrow component}
\tablenotetext{c}{Broad component}
\end{table}

\end{appendix}

\end{document}